\documentclass[twocolumn]{aastex631}

\begin{document}

\title{Constraining the Thickness of TRAPPIST-1 b's Atmosphere from its JWST Secondary Eclipse Observation at 15 $\mu$m}

\author[0000-0003-2775-653X]{Jegug Ih}
\affiliation{Department of Astronomy, University of Maryland, College Park, MD 20742, USA}

\author[0000-0002-1337-9051]{Eliza M.-R. Kempton}
\affiliation{Department of Astronomy, University of Maryland, College Park, MD 20742, USA}

\author[0000-0002-1518-7475]{Emily A. Whittaker}
\affiliation{University of California, Los Angeles, Department of Earth, Planetary, and Space Sciences, Los Angeles, CA 90095-1567, USA}

\author[0000-0001-9939-5564]{Madeline Lessard}
\affiliation{Department of Astronomy, University of Maryland, College Park, MD 20742, USA}



\begin{abstract}
Recently, the first JWST measurement of thermal emission from a rocky exoplanet was reported.  The inferred dayside brightness temperature of TRAPPIST-1 b at 15 $\mu$m is consistent with the planet having no atmosphere and therefore no mechanism by which to circulate heat to its nightside. In this Letter, we compare TRAPPIST-1 b’s measured secondary eclipse depth to predictions from a suite of self-consistent radiative-convective equilibrium models in order to quantify the maximum atmospheric thickness consistent with the observation.  We find that plausible atmospheres (i.e., those that contain at least 100 ppm CO$_2$) with {surface pressures greater than 0.3 bar are ruled out at 3$\sigma$}, regardless of the choice of background atmosphere, {and a Mars-like thin atmosphere with surface pressure 6.5 mbar composed entirely of CO$_2$ is also ruled out at 3$\sigma$}.  Thicker atmospheres of up to 10 bar (100 bar) are consistent with the data at 1$\sigma$ (3$\sigma$) only if the atmosphere lacks \textit{any} strong absorbers across the mid-IR wavelength range — a scenario that we deem unlikely.   We additionally model the emission spectra for bare-rock planets of various compositions.  {We find that a basaltic, metal-rich, and Fe-oxidized surface best matches the measured eclipse depth} to within 1$\sigma$, and the best-fit grey albedo is $0.02 \pm 0.11$.  We conclude that planned secondary eclipse observations at 12.8 $\mu$m will serve to validate TRAPPIST-1 b’s high observed brightness temperature, but are unlikely to further distinguish among the consistent atmospheric and bare-rock scenarios.
\end{abstract}



\section{Introduction} \label{sec:intro}

We have now entered the era of JWST, and with it comes the potential to perform the first meaningful characterization of terrestrial (i.e., rocky) exoplanets.  Among the possible rocky planet targets for JWST, those in the TRAPPIST-1 system are some of the most promising for atmospheric characterization due to their very favorable planet-to-star size ratios \citep{gillon16}.  The system is also of extreme interest because it hosts multiple terrestrial planets, including several that reside in or near the habitable zone \citep{gillon17}.  Recently, \citet{greene23} measured the thermal emission from the innermost planet, TRAPPIST-1~b, and found that its 15-$\mu$m brightness temperature is consistent with the planet being a bare rock, devoid of any atmosphere at all.  

Thermal emission measurements of presumed tidally-locked planets, such as those produced by \citet{greene23} for TRAPPIST-1 b, are a productive avenue for confirming whether rocky exoplanets possess atmospheres \citep{koll19,mansfield19}.  By measuring the planet’s dayside temperature via secondary eclipse observations, one can constrain the presence and thickness of the atmosphere in the following sense:  atmospheres serve to lower the dayside emission temperature below what would be expected for a bare (and dark) rocky surface.  Even moderately thick atmospheres transport considerable heat away from a tidally-locked planet’s dayside  \citep{koll22}.  Reflective aerosols, another signpost of a planet possessing an atmosphere, also serve to lower the dayside temperature by reflecting incoming stellar radiation back to space \citep{mansfield19}.  The maximal dayside effective temperature, corresponding to no atmosphere and a zero-albedo surface is: 
\begin{equation}
T_{max} = T_* \sqrt{\frac{R_*}{d}} \left( \frac{2}{3} \right) ^{1/4} 
\end{equation}
where $T_*$ and $R_*$ are the stellar {effective} temperature and radius, and $d$ is the planet-star separation.  For TRAPPIST-1 b, { $T_{max} = 508 \pm 6$~K,} whereas the 15 $\mu$m brightness temperature reported by Greene et al. is $503^{+26}_{-27}$ K, fully consistent with the no-atmosphere scenario.

From a theoretical standpoint, it is unclear whether terrestrial planets orbiting M-dwarfs should be expected to possess atmospheres.  There are studies that go both ways.  Atmospheric loss processes should be efficient for planets orbiting active M-dwarf host stars, but some planets may be able to retain their atmospheres or renew them via outgassing following a decline in stellar activity with age \citep[e.g.][]{zahnle17, turbet20, wordsworth22}.   

Observationally, to-date there are no studies that definitively confirm the presence of an atmosphere on a rocky exoplanet.  Flat transmission spectra are the norm \citep[e.g.][]{dewit18,diamondlowe18, diamondlowe20,mugnai21,libby22, lustig23}, and the few studies that have claimed detections of atmospheric spectral features for terrestrial exoplanets have been called into question or have ambiguous interpretation \citep[e.g.][]{southworth17,swain21,moran23}.  Thermal emission measurements of the planets LHS 3844b \citep{kreidberg19} and GJ 1252b \citep{crossfield22} have found dayside temperatures that are consistent with the no-atmosphere limit, the former by way of a full-orbit phase curve.  It stands to reason that less irradiated planets should be less susceptible to atmospheric loss, but TRAPPIST-1 b is the coldest planet yet to be subjected to the thermal emission test for possessing an atmosphere, yielding the same result of no apparent sign of a gaseous envelope.

In this Letter we quantify the range of atmospheres and surfaces that are consistent with the \citet{greene23} measurement of TRAPPIST-1 b’s secondary eclipse depth at 15 $\mu$m.  We show in what follows that thick atmospheres can be definitively ruled out by this single data point.  Given the range of scenarios that we still find to be consistent with the data, we also predict the degree to which further observations, including planned measurements at 12.8 $\mu$m, will be able to distinguish among the remaining plausible atmospheres and surfaces.







\section{Methods} \label{sec:method}

In this section, we describe our model and parameter choices.  To calculate the eclipse spectrum of different surfaces and atmospheres, we use \textsc{helios}, an open-source 1D radiative transfer code that computes the thermal profile of a planetary atmosphere in radiative-convective equilibrium \citep{malik17,malik19,malik19b}.  Most of our approach closely follows \citet{whittaker22}, which performed a similar analysis for the \textit{Spitzer} observation of LHS 3844 b, and we refer the readers to that work for more details of the modelling.

One key detail worth mentioning here is that we calculate the heat redistribution factor ($f$) self-consistently with the radiative transfer using the analytical approximation in \citet[equation 10]{koll22}.  In the approximation, $f$ depends on the equilibrium temperature, the surface pressure, and the longwave optical depth at the surface; \textsc{helios} has the ability to iterate to a value of $f$ that satisfies global energy balance.  We note a caveat that this method subtracts the approximated transported heat from the incident stellar flux to calculate the dayside energy budget, but does not consider the vertical dependence of the day-to-night heat flow; hence the redistribution could be construed to happen either uniformly or at the top of the atmosphere in our models.

We model a range of surface pressures that is broad enough span full redistribution ($f=1/4$) to no redistribution ($f=2/3$), resulting in a surface pressure grid of $10^{-4}$ bars to $10^2$ bars, spaced at 1 dex.  For the composition of the atmospheres, in addition to a 100\% CO$_2$ atmosphere, we choose to vary the abundance of trace CO$_2$, at 1 ppm, 100 ppm, and 1\%, against background gases of N$_2$, O$_2$, and H$_2$O.  Moreover, we also consider atmospheres containing a range of other trace gases plausible in secondary atmospheres \citep[][]{turbet20, krissansen21, whittaker22}, which may not necessarily absorb at 15 $\mu$m but may be detected via observations at other wavelengths. For this purpose, we adopt the same trace abundance grids (i.e.\ 1 ppm, 100 ppm, 1\%) for CO, CH$_4$, H$_2$O, and SO$_2$, against a background gas of N$_2$ for the former two and O$_2$ for the latter.  SO$_2$ is unique in that it has broad infrared absorption features just outside the 15-$\mu$m bandpass, which produce interesting implications for observations at 15 $\mu$m; we discuss this further in Section~\ref{sec:results}.  For all models, we assume an {intrinsic temperature} of $T_{\mathrm{int}}=0 \mathrm{K}$.


For all of the atmosphere models, we adopt a surface albedo of 0 (i.e.\ a true blackbody), to produce the maximum limit on the atmospheric pressure consistent with the observation; any value of non-zero albedo will dilute the energy budget and decrease the eclipse depth, thereby making a model at a given atmospheric pressure even less consistent with the observation. 

Given that TRAPPIST-1 b's dayside temperature is consistent with the no-atmosphere limit, we also explore a number of bare surface models that have no atmospheres at all.  Here the eclipse spectrum instead arises due to the wavelength-dependent albedo spectrum of the surfaces.  We consider {six} surfaces that are plausible, given the level of irradiation received by TRAPPIST-1 b: basaltic, ultramafic, feldspathic, {metal-rich, Fe-oxidized,} and granitoid \citep{hu12, mansfield19}.  We also run a number of grey albedo surfaces at $A = 0.2, 0.4, 0.6, 0.8, 0.95$.


We adopt the stellar and planetary parameters as obtained in \citet{agol21}.  We use the \textsc{sphinx} stellar model spectrum grid \citep{iyer23} interpolated to TRAPPIST-1 parameters assuming solar composition to calculate the thermal profile and the eclipse depth of the planet.  \textsc{sphinx} models are expected to better model the stellar spectra at such low temperature ranges than the typical \textsc{phoenix} models, using updated line lists \citep{iyer23}.  Indeed, we find that the \textsc{sphinx} model reproduces the observed stellar flux at 15 $\mu$m better than the \textsc{phoenix} model \citep[to within 7\% versus 13\%; see Methods of][]{greene23}

After obtaining the eclipse spectra, we calculate the binned depth at the photometric band of F1500W; {we integrate the planetary flux weighted by the bandpass function, then integrate the stellar flux weighted by the same function, and then obtain the ratio of the two.}  We perform the same calculation for F1280W to make predictions for upcoming observations.  The F1280W bandpass lies outside the CO$_2$ absorption feature, and the difference between the two bandpasses serves as a metric to constrain either atmospheric pressure, CO$_2$ abundance, or both \citep{deming09}.

We calculate the brightness temperature ($T_{\mathrm{b}}$) in the F1500W filter by determining the temperature of the blackbody whose eclipse depth (obtained via identical weighting and integrating as for the planetary flux) matches the observed eclipse depth.  {We note that this calculation differs slightly from the procedure followed by \citet{greene23}, who found the temperature of the blackbody whose per-frequency flux evaluated at the ``effective" filter wavelength matched the observed per-frequency planetary flux.  Our calculation leads to a best-fit brightness temperature of $T_{\mathrm{b}}=505 \pm 27$K, rather than the $T_{\mathrm{b}}=503^{+26}_{-27}$K reported in \citet{greene23}.  Given the uncertainty, this minor discrepancy will not impact our analysis.}

\section{Results} \label{sec:results}

\begin{figure*}[t!]
    \centering
    \includegraphics[width=\textwidth]{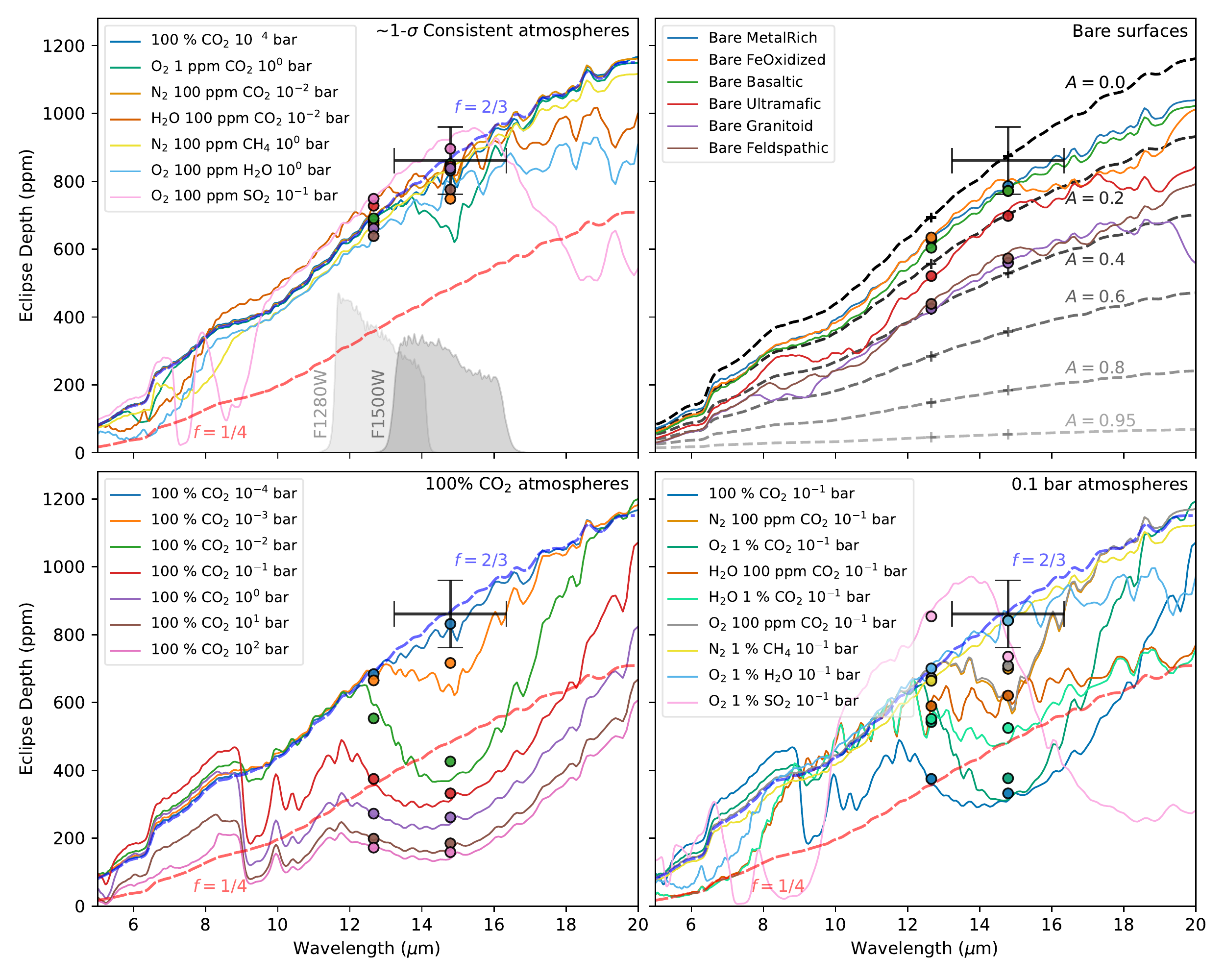}
    \caption{The eclipse spectra of various models run in this study.  We show: a suite of atmospheric models that are 1-$\sigma$ consistent with the observation (\textit{top left}); bare surface models, which are all consistent with the observation (\textit{top right}); 100 \% CO$_2$ atmosphere models at various surface pressures (\textit{bottom left}); and models with surface pressures of 0.1 bar, varying the compositions (\textit{bottom right}).  The compositions denote that the first species is the dominant species, with the second species in indicated trace amounts.  The binned depths at F1500W and F1280W are shown as markers, as well as each bandpass function weighted by the stellar spectrum.  We also show, in dashed lines, the eclipse depths resulting from blackbodies at 508 K (blue) and 400 K (red), corresponding to no redistribution ($f=2/3$) and full redistribution ($f=1/4$), respectively. On the upper right panel, dashed lines indicate grey albedo surface models.  {The features in the blackbody eclipse spectrum arise due to spectral features in the \textit{stellar} spectrum.}}
    \label{fig:spectra}
\end{figure*}

\begin{figure*}[t!]
    \centering
    \includegraphics[width=\textwidth]{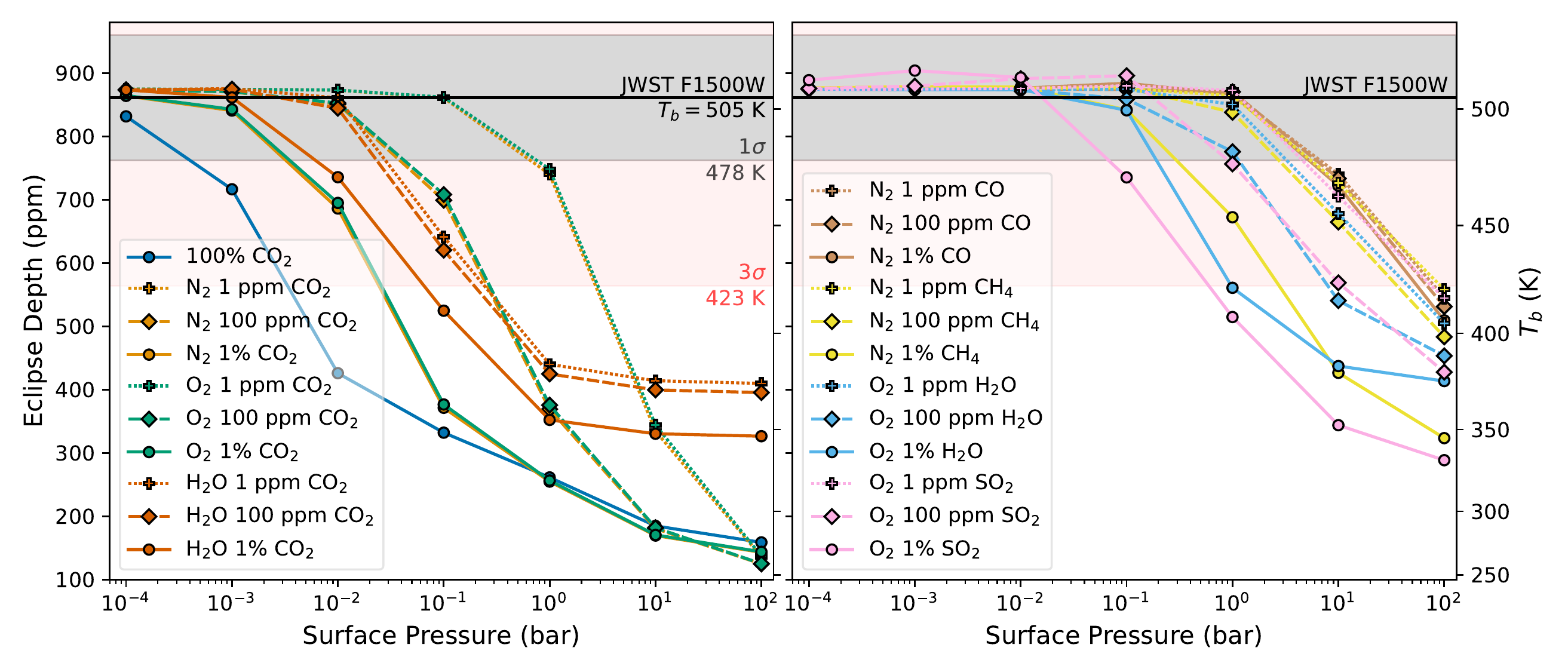}
    \caption{The binned eclipse depths and their brightness temperature in the F1500W band for all of the atmospheric models run, varying the pressure of the atmosphere at the surface.  Models atmospheres that do and do not include CO$_2$ are shown in the left and the right panel, respectively.  The measured eclipse depth from \citet{greene23} is shown as the solid black line, and its 1-$\sigma$ (grey) and 3-$\sigma$ (red) uncertainties are are also shown, as well as the corresponding brightness temperatures.  The compositions denote that the first species is the dominant species, with the second species in indicated trace amounts.  Atmospheres with $\geq$100 ppm CO$_2$ are consistent with the measurement at 1$\sigma$ only if the atmospheric pressure is less than 0.1 bar.} 
    \label{fig:depths}
\end{figure*}

\begin{figure*}[t!]
    \centering
    \includegraphics[width=\textwidth]{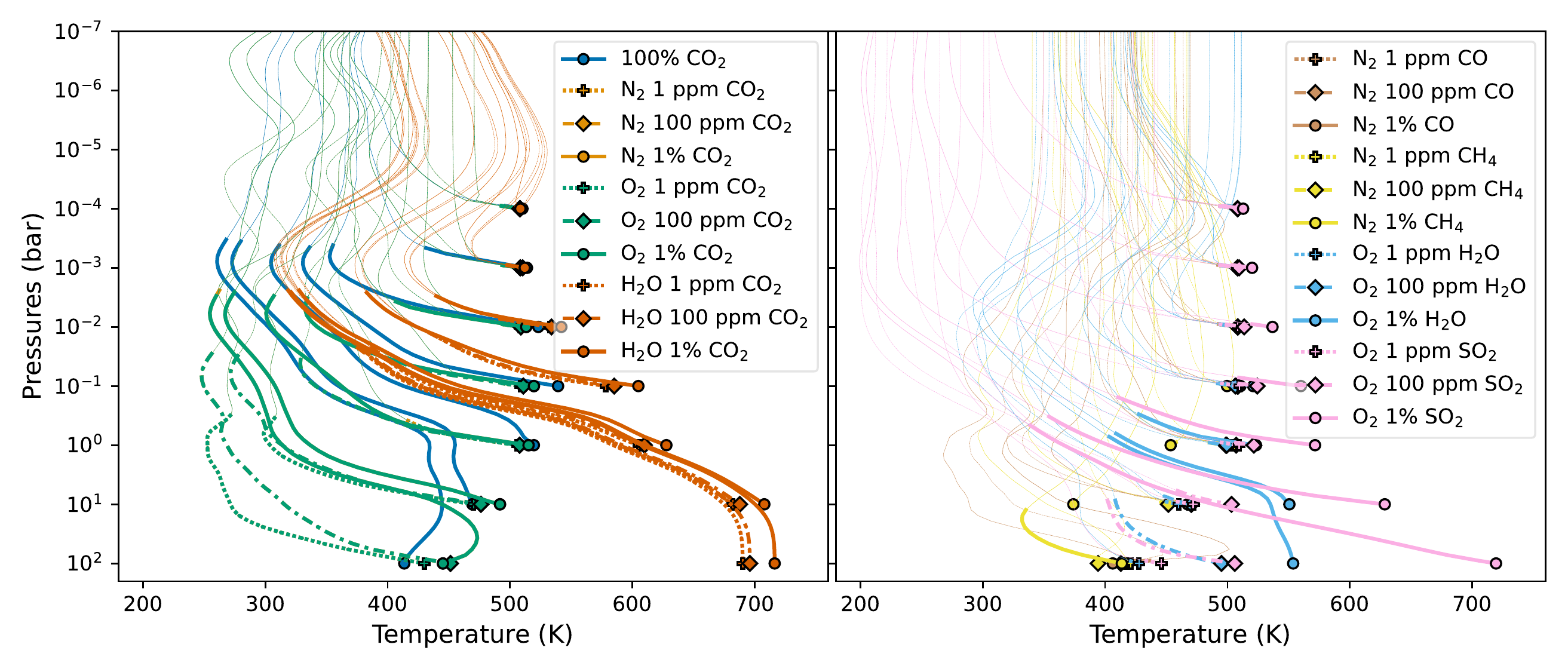}
    \caption{The temperature-pressure (T-P) profiles of the model atmospheres in radiative-convective equilibrium.  Models atmospheres that do and do not include CO$_2$ are shown in the left and the right panel, respectively, similarly to Figure \ref{fig:depths}.  The optically thick region of the T-P profiles below the photosphere ($\tau=2/3$) at $\lambda=14.79~\mu$m are shown with thick lines.  The markers indicate the surface pressure of each model atmosphere.  The compositions denote that the first species is the dominant species, with the second species in indicated trace amounts.  The N$_2$ and O$_2$-dominated atmospheres completely overlap in the left panel.  It can be seen that while near-infrared absorbers such as H$_2$O can cause thermal inversions, they occur at regions where the atmosphere is optically thin and hence will not result in emission features in the spectra.  For most of the models that do not contain CO$_2$, the atmosphere is optically thin in the F1500W bandpass down to the surface.}
     \label{fig:tp}
     \vspace{0.6cm}
\end{figure*}


\subsection{Atmospheric Thickness and Surface Composition}

Our results support the general conclusion from \citet{greene23} that TRAPPIST-1 b  does not possess a thick atmosphere. We will present the maximum atmospheric thickness consistent with the observed eclipse depth of $861 \pm 99$ ppm for each set of model composition and also highlight  interesting behaviors from a theoretical perspective.  We show the eclipse spectra for selected atmospheric and surface models in Figure \ref{fig:spectra} and the binned eclipse depths for all of the atmospheric models in Figure \ref{fig:depths}, varying the composition and the surface pressure.  The accompanying temperature-pressure (T-P) profiles for each of the atmosphere models are shown in Figure~\ref{fig:tp}.

\subsubsection{Atmospheres with CO$_2$}

We posit that TRAPPIST-1 b should realistically have at least moderate amounts of CO$_2$ if it does possess an atmosphere.  This statement is in line with theoretical studies of the atmosphere of TRAPPIST-1 b and in general of rocky exoplanets receiving a comparable degree of irradiation \citep{lincowski18, hu20, turbet20}.  CO$_2$ is robustly expected to be present in non-hydrogen-dominated atmospheres \citep[e.g., as indicated for TRAPPIST-1 b from its transmission spectrum;][]{dewit16}, and the gas is robust against various escape processes, although photodissociation can deplete its abundance.


Pure CO$_2$ atmospheres are 1-$\sigma$ consistent with the eclipse measurement for surface pressures up to 0.4 mbar and 3-$\sigma$ consistent up to 3 mbar (Figure \ref{fig:depths}), indicating that even a Mars-like thin atmosphere ($P_{\mathrm{surf}} = 6.5$ mbar) composed entirely of CO$_2$ is unambiguously ruled out. To first order, the secondary eclipse depth depends on the \textit{partial pressure} of CO$_2$, so the atmosphere may be thicker if the CO$_2$ abundance (i.e.\ its mixing ratio) is smaller.  N$_2$ or O$_2$-dominated atmospheres with $\geq$100 ppm of CO$_2$ are 1-$\sigma$ consistent at 0.04 bar at most, and 1 bar atmospheres are ruled out by more than 3$\sigma$.



The presence of H$_2$O has a non-trivial effect on the eclipse spectrum as it both increases the absorption and changes the thermal structure.  For instance, at a surface pressure of 0.1 bar, H$_2$O-dominated atmospheres with 1 ppm or 100 ppm CO$_2$ have deeper eclipse depths than the corresponding O$_2$ or N$_2$-dominated atmospheres, while the one with 1\% CO$_2$ has a shallower depth than atmospheres with the other background gases.  Additionally, the lower atmosphere becomes much hotter for the thicker H$_2$O-dominated atmospheres due to greenhouse heating being more effective than the cooling of day-night redistribution.

\begin{figure*}[t!]
    \includegraphics[width=\textwidth]{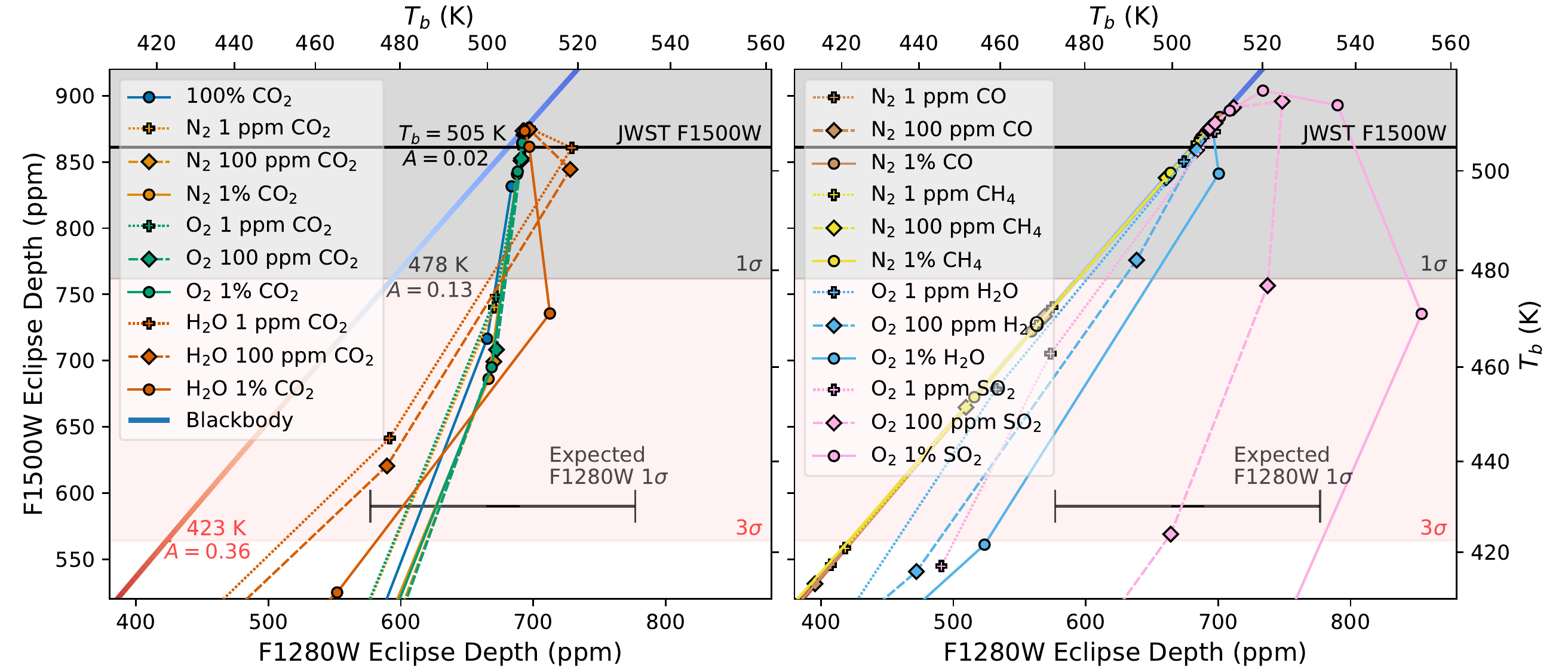}
    \caption{A {color-color-like diagram of} predicted binned eclipse depths in the F1280W band (horizontal axis) and the binned F1500W eclipse depths for all of the model atmospheres, along with their brightness temperatures ($T_{\mathrm{b}}$) in each band.   Models atmospheres that do and do not include CO$_2$ are shown in the left and the right panel, respectively.    The measured eclipse depth from \citet{greene23} is shown as the solid black line, and its 1-$\sigma$ (grey) and 3-$\sigma$ (red) uncertainties are are also shown.  The vertical axis is identical to Figure \ref{fig:depths}, but is zoomed to focus on models consistent with the F1500W observation, alongside the expected F1280W uncertainty ($\sim 100$ ppm) shown as an errorbar. {The binned eclipse depths for a blackbody over a range of temperatures is shown as a multi-colored line. The temperature of the blackbody can be read off from the $T_{\mathrm{b}}$ in either axes, by definition.} The corresponding $T_{\mathrm{b}}$ and bond albedo ($A$) at each confidence interval is also shown.  {All models that include CO$_2$ (in the left panel) lie on the right side of the blackbody line, indicating a higher $T_{\mathrm{b}}$ in the F1280W than in F1500W due to the CO$_2$ \textit{absorption} at 15 micron.}  The compositions denote that the first species is the dominant species, with the second species in indicated trace amounts.  As one follows each composition line, atmospheric pressure starts at $10^{-4}$ bar close to the observed F1500W measurement and increases in 1-dex intervals as in Figure \ref{fig:depths} with generally decreasing 15-$\mu$m eclipse depths.  We do not show the bare surface depths in this figure, but they lie close to the blackbody line and deviate less than 25 ppm in either bandpass.}
     \label{fig:1280}
\end{figure*}

H$_2$O is also interesting in that it can generate thermal inversions in planets orbiting M stars \citep{malik19}.  Thermal inversions are interesting in the context of the \citet{greene23} secondary eclipse measurement because they have the potential to reverse absorption features into emission, opening a possibility that the high observed 15-$\mu$m brightness temperature could be due to a CO$_2$ \textit{emission} feature originating from from a thick(er) atmosphere. For TRAPPIST-1 b, we indeed find that H$_2$O causes thermal inversions (Figure \ref{fig:tp}), but they occur in the upper atmosphere well above the IR photosphere and thus do not significantly impact the shape of the 15-$\mu$m CO$_2$ feature, which uniformly appears in absorption in all of the models we have produced.  We have also experimented with different mixtures of O$_2$, H$_2$O, and CO$_2$ (not shown), but find that no combination leads to emission features.  {In fact, in Figure \ref{fig:1280}, one can see that the brightness temperature at 15 micron is lower than that at 12.8 micron for every model, indicating CO$_2$ absorption, rather than emission, is being observed.}




\subsubsection{Atmospheres with no CO$_2$}

While less plausible chemically, atmospheres that do not contain any CO$_2$ at all remain consistent with the secondary eclipse measurement to higher surface pressures.  Atmospheres that have CO or CH$_4$ as the trace gas are 1-$\sigma$ consistent to 1 bar for all trace abundances, except the 1\% CH$_4$ model which has a shallower depth that is 2-$\sigma$ consistent.  In Figure \ref{fig:tp}, it can be seen in the right panel that all of these atmospheres except the 1\% CH$_4$ 10$^2$ bar model remain optically thin in the 15~$\mu$m bandpass down to the surface, and the change in eclipse depth with surface pressure is due to the cooling effect of redistribution.  Atmospheres with trace H$_2$O behave similarly except that the 1\% H$_2$O atmospheres becomes optically thick at atmospheric pressures around 0.1 bar, and the eclipse depth is already $>$ 3-$\sigma$ inconsistent for a surface pressure of 1 bar.

Atmospheres with trace SO$_2$ behave somewhat differently since SO$_2$ has a broad absorption feature at wavelengths just redward of the 15-$\mu$m bandpass.  For moderate SO$_2$ abundances (e.g.\ the pink line for the 100 ppm 0.1 bar atmosphere in the top left panel of Figure \ref{fig:spectra}), the strong absorption at $\sim$18--20 $\mu$m pushes more flux into the 15-$\mu$m bandpass, leading to \textit{increased} planetary emission over the wavelength range of the \citet{greene23} secondary eclipse observation.  The emission from a transparent spectral window is therefore a plausible mechanism for increasing the secondary eclipse depth in a single bandpass, but it comes at the cost of sharply reduced fluxes at other wavelengths; this effect can therefore be diagnosed with additional spectroscopic observations. For higher SO$_2$ abundances however, the absorption feature is strong enough to affect the 15-$\mu$m bandpass, and it therefore has the opposite effect of reducing the eclipse depth in the F1500W filter (Figure \ref{fig:spectra}, {pink} line in bottom right panel).  This indicates that the nature of the absorber needs to be very finely tuned to match the \citet{greene23} measurement.  


\subsubsection{Bare surfaces}


If TRAPPIST-1 b truly has no atmosphere whatsoever, {we find that the F1500W measurement is consistent with a bare rock planet with a basaltic, Fe-oxidized, or metal-rich surface to within 1$\sigma$, while granitoid and feldspathic surfaces are ruled out at more than 3$\sigma$ (Figure~\ref{fig:spectra}, top right panel).  The latter two materials have high albedos around 1 $\mu$m where the luminosity of the TRAPPIST-1 host star is greatest \citep{hu12,mansfield19}, thus reducing the energy received by the planet and lowering the temperature at which it radiates.  The fact that we can rule out some surface compositions demonstrates the utility of secondary eclipse spectroscopy for constraining the surface properties of rocky exoplanets.  However, \citet{mansfield19} point out that granitoid and feldspathic surfaces (the ones that we rule out here) are also among those that are implausible for hot rocky planets like TRAPPIST-1 b, as they either require liquid water to form or they are unlikely to be able to form on larger planets \citep{elkins12}.}  Among grey surfaces, we find that the best-fit surface albedo is $0.02 \pm 0.11$.


\subsection{Prospects for Future Observations}

Given the various atmospheres and surfaces that remain consistent with the \citet{greene23} 15 $\mu$m secondary eclipse measurement, we investigate here the possibility that additional observations could help to further constrain the properties of TRAPPIST-1 b.  In particular, five secondary eclipses are slated to be observed with MIRI F1280W filter centered on 12.8 $\mu$m to provide a second spectroscopic data point for \mbox{TRAPPIST-1 b's} thermal emission.  In Figure~\ref{fig:1280} we show the eclipse depths from our models binned to the F1280W bandpass against the the binned eclipse depth in the F1500W bandpass.

The F1280W is intended to observe the eclipse depth out of the CO$_2$ band such that the difference between the two provides a constraint on the atmospheric pressure and possibly composition, but the very high eclipse depth of F1500W alone already provides a firm constraint on the brightness temperature and hence the atmospheric pressure.  Assuming an observation uncertainty comparable to that of F1500W (99 ppm), the F1280W secondary eclipse is unlikely to help further distinguish between, for example, a very thin 10$^{-4}$ bar 100\% CO$_2$ atmosphere, a 1 bar O$_2$-dominated 1 ppm CO$_2$ atmosphere, a 1 bar N$_2$-dominated atmosphere with 100 ppm CH$_4$  as they all fall roughly within a span of 100 ppm.  Therefore, we conclude that the F1280W observation will be most useful for validating the high brightness temperature of TRAPPIST-1 b as observed by F1500W.


Indeed, in Figure \ref{fig:spectra}, most 1-$\sigma$ consistent spectra follow the $f=2/3$ blackbody spectrum (blue dashed line) closely down to 10 $\mu$m, and only at shorter wavelengths do spectroscopic absorption features appear.  However, due to the small eclipse depth at these wavelengths, spectroscopy using MIRI LRS with nominal uncertainty of (say) 30 ppm at a spectral resolution of $R=10$ will be able to distinguish only between end-member cases at best rather than tightly constraining the composition and the surface pressure.  Namely, if the planet has H$_2$O, CH$_4$, or SO$_2$, absorption features between 5--10 $\mu$m, MIRI LRS could be used to distinguish between an airless blackbody and a thin atmosphere.

As for distinguishing among bare rock surfaces, the additional F1280W observation is unlikely to be helpful for this purpose as the binned eclipse depths {of consistent surfaces} are very similar (Figure \ref{fig:spectra}).  The surfaces are generally difficult to distinguish across all wavelengths that MIRI can observe in.  









\section{Discussion and Summary} \label{summary}


We have shown that, based on the \citet{greene23} secondary eclipse observation at 15 $\mu$m, TRAPPIST-1 b does not appear to host a thick atmosphere.  Formally, our models rule out atmospheres with at least 100 ppm CO$_2$ thicker than 0.3 bars at 3$\sigma$.  For a 100\% CO$_2$ atmosphere (i.e., a Mars or Venus-like composition), the atmosphere must be less than 3 mbar thick at 3$\sigma$ confidence {to be consistent with the measured eclipse depth at 15 $\mu$m}.  We argue that TRAPPIST-1 b is unlikely to host an atmosphere devoid of CO$_2$, and therefore atmospheres thicker than $\sim$0.1 bar are ruled out.  {Various types of geophysically plausible rocky surfaces are all consistent with the \citet{greene23} measurement, and the eclipse observation rules out less plausible granitoid and feldspathic surfaces.} The best-fit grey surface albedo is $0.02 \pm 0.11$.

The 1-$\sigma$ consistent atmospheres and surfaces that we identify in this Letter will be difficult to distinguish with upcoming JWST observations except perhaps the very end-member scenarios.  The predicted eclipse depths for the F1280W filter are close enough to each other to be within the uncertainty of the observation.  MIRI LRS may be able to distinguish between a bare rock and a 0.1 bar H$_2$O-dominated atmosphere by measuring the eclipse spectrum from 5-10 $\mu$m, but there are many degenerate scenarios in between.  Finally, the planned NIRISS SOSS observation of TRAPPIST-1 b via complementary measurements in \textit{transmission} \citep[Cycle 1 GO 2589]{lim21} also aims to distinguish between a bare rock and a thin atmosphere. {In the case of a clear atmosphere, transmission spectroscopy can generally provide a signal that is easier to interpret than that of thermal emission, since H$_2$O and CO$_2$ features should be detectable.  Transmission spectroscopy is also more agnostic to the thermal structure of the atmosphere and could therefore provide a less ambiguous constraint on the composition.  On the other hand, } transmission spectroscopy of small, rocky planets is challenging as the high mean molecular weight of secondary atmospheres {and aerosols (if present)} render the transmission spectrum closer to a flat spectrum, which is indistinguishable from a bare rock planet \citep{millerricci09, barstow16, ducrot20}.  Additionally, host stellar effects also leave an imprint on the transmission spectrum, leading to spectral contamination that can be difficult to disentangle from \textit{bona fide} atmospheric features \citep{rackham18, rachkam23, moran23}.

We have neglected the radiative effects of clouds in our work.  The clear atmosphere T-P profiles in Figure~\ref{fig:tp} do cross condensation curves such that water or sulfur clouds can form \citep{mbarek16, lincowski18}.  However, clouds of appreciable column density will have higher albedos than rocky surfaces \citep[Fig. 6]{mansfield19} and are inconsistent with the observation, given such a low inferred albedo (even with the uncertainties taken into account).  Additionally, climate modelling suggests that aerosols are unlikely to form in TRAPPIST-1 b \citep{lincowski18}.  As such, we find the scenario that the planet hosts an atmosphere with a reflecting cloud to be inconsistent with the \citet{greene23} secondary eclipse measurement.


The F1500W observations of TRAPPIST-1 b demonstrate the utility of {secondary eclipse observations} for determining whether rocky planets possess atmospheres {and for constraining their surface composition.}  Secondary eclipse observations will soon also be applied to other rocky planets around M dwarfs, with observation planned for more targets such as TRAPPIST-1 c \citep[Cycle 1 GO 2304]{kreidberg21a},  Gl 486 b \citep[Cycle 1 GO 1743]{mansfield21}, GJ 1132 b \citep[Cycle 1 GTO]{lunine17}, and LHS 3844 b \citep[Cycle 1 GO 1846]{kreidberg21b}.  { The latter three use MIRI LRS rather than F1500W; an identical analysis to the current work can be performed by binning the entire 8--12 $\mu$m LRS spectrum to create a single broad photometric bandpass \citep[see e.g. \S3 of][]{koll19}, and the additional \textit{spectral} information can be used to further constrain the composition of the atmosphere or the surface \citep{whittaker22}.}  A larger sample of rocky planet targets observed in secondary eclipse will also help to answer population-level questions of whether rocky planets around M dwarfs can really host atmospheres and identify the ideal parameter space for establishing regimes in which they can.











\begin{acknowledgments}
We thank Tom Greene for useful discussion and for providing additional context on his JWST observations.  JI and EMRK acknowledge funding from the Alfred P. Sloan Foundation under grant G202114194.  {We thank the anonymous referee for useful feedback that helped to improve this manuscript.} 
\end{acknowledgments}


\bibliography{paper}{}
\bibliographystyle{aasjournal}



\end{document}